# Predicting the Growth of Two-Dimensional Nanostructures

B. Viswanath, Paromita Kundu, B. Mukherjee and N. Ravishankar*

*Materials Research Center, Indian Institute of Science, Bangalore 560012 India*

**Abstract:**

**The ability to predict the morphology of crystals formed by chemical reactions is of fundamental importance for the shape-controlled synthesis of nanostructures. Based on the atomistic mechanism for crystal growth under different driving forces, we have developed morphology diagrams to predict regimes for the growth of two-dimensional crystals. By using controlled reactions for crystal growth in the absence of surfactants/capping agents, we demonstrate the validity of this approach for the formation of 2-D structures of Au, Ag, Pt, Pd and hydroxyapatite.**

An understanding of the external morphology of crystals and its relation to internal structure has been a topic of active study since the times of Kepler. It is well-recognised that the morphology of nanostructures affects their properties profoundly [1-8]. Two-dimensional nanostructures in the form of platelets/sheets, nanoprisms and belts exhibit intriguing properties that have several potential applications [1-4,6,7,9-11]. In spite of the availability of a number of methods [2-8,12-15], the mechanism of formation of such structures remains elusive. Mechanisms based on preferential adsorption of surfactants [8,16], oriented attachment [16], soft templates [17,18], aggregation of spherical particles [4] and kinetic control [6-8,18,19] have been proposed in the literature but do not satisfactorily explain the shape control. Here, we show for the first time that this problem can be analysed based on classical crystal growth concepts. Based on the atomistic mechanism for crystal growth under different driving forces [20,21], we have developed morphology diagrams to predict conditions under which two-dimensional crystals form. By using controlled reactions for crystal growth in the absence of reducing and capping agents or by precipitation under controlled conditions coupled with detailed microstructural evidence, we demonstrate the validity of this approach for the formation of 2-D nanostructures of Au, Ag, Pt, Pd and hydroxyapatite. The analysis and experiments have important consequences for rational synthesis of two-dimensional nanostructures and answer some long-standing questions related to their growth. The generality of the analysis implies that it can be used to predict regimes of two-dimensional growth in a variety of systems ranging from crystals synthesized from a solution phase, from the vapor phase and inorganic phases formed by biomineralization.

The formation of crystals from vapor or in a liquid phase proceeds by nucleation and growth mechanism with the driving force given by the associated volume free energy change. Based on the driving force, two distinct growth regimes can be identified. At large driving forces, the interface can move normal to itself leading to a continuous growth (Fig. 1A). At low driving forces, however, growth has to rely on the formation of steps and a lateral motion of steps on the surface [20-22]. It has been proposed that screw dislocations enable crystal growth to proceed at very low driving forces by providing a constant supply of kink sites at the surface [22]. However, in the absence of screw dislocations, growth has to proceed by nucleation of two-dimensional islands at the growing interface (Fig. 1B). It has been shown that the critical driving force for the continuous growth to take place is given as $-\Delta G > \pi\sigma g/a$, where '$\sigma$' is the interfacial/surface free energy depending on the medium where the crystal is forming, '$g$' is a measure of the diffuseness of the interface (taken to be 1 for sharp interfaces) and '$a$' is the monatomic step height on the surface [20,21]. Also for $-\Delta G < \sigma g/a$, growth has to proceed by the two-dimensional nucleation mechanism involving lateral motion of steps. The range $\sigma g/a < -\Delta G < \pi\sigma g/a$ represents the transition regime where two-dimensional nucleation takes place at lower driving forces with a gradual transition to continuous growth at larger driving forces [20,21]. In the context of solidification in a one-component system for which the original theory was developed, the degree of undercooling is the only tunable parameter for varying the driving force. On the other hand, for chemical reactions, there are several parameters like pH, temperature and concentration of reactants that can be varied to tune the driving force over a large range of values. As an extreme case, the reaction can be made to progress in the backward direction in which case the free energy change for the forward reaction becomes positive. The driving force can be quantified by calculating the free energy for the reaction and thus regimes where different mechanisms will be operative can be identified (See Supplementary Information). To the best of our knowledge, this is the first time that such an analysis has been extended for products formed as a result of chemical reactions. The morphology diagrams, thus developed, enable rational synthesis of shape-controlled two-dimensional nanostructures.

In reported wet-chemical synthesis for two-dimensional structures, the presence of a large number of reagents including reducing agents and capping agents complicates the interpretation of the mechanism of shape control. Hence, we follow a procedure where external reducing agent or capping agent is not used. The reaction involving the oxidation of water has been used to reduce the noble metal salts to the metallic state in aqueous medium. The free energy change can be calculated as a function of concentration of the reactants, pH and the temperature of the reaction (See Supplementary Information). Based on the interfacial energies and monatomic step heights, the temperature and pH regimes for 2-D growth and 3-D growth can be identified based on whether $-\Delta G < \sigma/a$ or $-\Delta G > \pi\sigma/a$ respectively. Fig. 1C-F illustrates the resulting morphology diagrams for the case of Au, Pt, Ag and Pd from such an analysis clearly delineating the regimes where the 2-D and 3-D morphologies can be observed. We define the 2-D structures as the ones where one dimension (thickness) is much smaller than the other two dimensions and 3-D structures as equiaxed structures.

The grey regions in the diagram represent regions where the free energy change for the forward reaction is positive and hence there will be no reduction in that regime. An increase in pH or temperature causes a change in sign of the free energy and one can obtain 2-D structures in the regions marked yellow. Above a critical driving force, the interface can advance normal to itself and 3-D structures are expected to form (red region). The intermediate regime (marked green) represents the transition zone from the 2-D to 3-D regime with the barrier for step nucleation gradually vanishing as one approaches the 3-D regime. The experimental data points are also represented in the same diagram (triangles representing conditions under which 2-D structures formed and circles representing conditions under which 3-D structures formed). The experimental observation is consistent with the predictions of the morphology diagrams illustrated here. It is to be noted that the driving force for chemical reactions changes as the reaction proceeds (owing to changes in pH and concentration). The use of a buffer allows the pH to be maintained in the course of the reaction. The experimental points represent the driving force at the start of the reaction. We have ensured that the change in driving force during the course of the reaction is not large enough to cause a change in the mechanism of growth. The main objective has been to illustrate the applicability of the general principles for a variety of systems rather than populate all the regions of the morphology diagram and thus we have only represented a limited number of experimental points in each case. In addition, the diagrams have been plotted under the assumption that the reaction for which $\Delta G$ is calculated takes place for all the pH and temperature ranges. However, this is often not the case. For instance, Pd tends to form hydroxides at pH above 4 and thus will not form under those conditions.

The morphology diagrams illustrated above predict that heating aqueous solutions of noble metal salts under suitable temperature and pH conditions should result in the formation of 2-D structures. Fig. 2A-D shows that this is indeed the case and illustrates TEM images of two-dimensional plate structures formed in the case of Au, Pt, Ag and Pd, respectively. The necessary and sufficient condition for the formation of 2-D structures is that the broad faces grow by the 2-D nucleation mechanism and that the growth rate of the side facets is much higher than the flat facets. It is to be noted that no external surfactant has been added to achieve this shape control. The plates, thus formed are very thin (tens of nanometres) as is evident from the fact that they are electron transparent and exhibit bend contours characteristic of thin metal platelets. Selected-area diffraction pattern from the platelets confirms the [111] orientation of the platelets. In all the systems, most of the platelets exhibit triangular/hexagonal shapes while some exhibit zigzagged morphologies that span several microns with the common feature being that the edges of the platelets typically run along a <110> direction. Increasing the driving force for the reactions (by increasing the pH or the temperature) results in the formation of extended 3-D structures as is seen from the SEM images in Fig. 2E-H.

A careful analysis of the defects present in the 2-D nanostructures provides useful clues about their mechanism of formation. Growth of FCC metals by the 2-D nucleation mechanism proceed by the nucleation of a single layer of (111) on a pre-existing (111) surface. Fig. 3A is an SEM image of a Pt platelet clearly showing the presence of steps on the surface. The platelet thickens by the lateral motion of these steps. Fig. 3B is a differential interference contrast image of gold platelets showing the presence of steps on the surface. Of course, the steps illustrated here are several atomic layers thick formed due to the bunching of monatomic steps on the surface. These clearly indicate that the plates thicken by the lateral motion of steps on the surface (growth by the two-dimensional nucleation mechanism). The defects in the platelets are initiated due to the nucleation of a 2-D layer that differs in orientation from the bulk crystal. In the case of Au and Ag platelets, formation of stacking faults has been reported [23]. The nucleation of a 'C' layer on an 'A' layer leads to the formation of …ABCACABC… type stacking viz., the formation of a stacking fault. The energy difference between the perfect crystal and a crystal containing a stacking fault is very small for Ag and Au (~ 20 $mJ/m^2$) and this explains the fact that stacking faults are often seen in the plates (as is evident from the presence of the kinematically forbidden 1/3(422) reflection in the SAED pattern (Fig. 3C) [23]. The fringes corresponding to 1/3 {422} are also seen in the high resolution image shown in Fig. 3D.

The stacking fault energy of Pt is very high (~ 373 mJ/m$^2$) [24] and hence stacking faults are not observed in the case of Pt (absence of the 1/3 (422) reflections). However, it is interesting to note that twist boundaries are often seen. Fig. 3E is an SAED pattern obtained from a Pt platelet. The diffraction pattern comprises two <111> zone patterns rotated by 27.8$^o$ about the common <111> axis. The inner ring of spots arises due to double diffraction as illustrated in the schematic in Fig.3F. There is a marked preference for the formation of the (111) twist boundary (Σ13) that is formed by a rotation of 27.8$^o$ about the <111> axis as was evident from the patterns obtained from several different platelets. A bright-field image from a platelet containing this boundary is shown in Fig. 3G. Indexing of the bend contours using dark-field imaging clearly shows that the alternate bend contours arise from the two different orientations of the platelets. This is strong evidence that this is in fact a twist boundary rather than two crystals lying one on the top of the other. Although, there are other twist orientations for a bulk crystal that have a lower energy, it is likely that relaxation that can take place in a single nucleating layer on a surface may stabilise this orientation over other orientations. Fig. 3H is a schematic illustrating an un-relaxed Σ13 boundary that is formed by rotating two crystals by 27.8$^o$ about a <111> axis. The observation of these defects clearly supports the two-dimensional nucleation mechanism of crystal growth by successive addition of (111) layers.

The applicability of the analysis has also been tested for a case where a common reducing agent (sodium borohydride) has been used for the reduction of silver. The calculation of the driving force become more complicated as the external reducing agent $NaBH_4$ involved in the reaction. Figure 4A shows the developed morphology diagram for the Ag-$NaBH_4$ case and the experiments carried out in 2D-3D transition regime resulted in 2D shapes whereas and reactions at 3D regimes produced 3D shapes of Ag and falls within the prediction of the morphology diagram. The exact 2D regimes for Ag cannot be obtained in the presence of strong reducing agent $NaBH_4$ as it required more acidic pH beyond the limit. Figure 4B shows the representative TEM-bright field image of Ag platelet showing the bend contours that are corresponds to the 111 plane.

The morphology diagrams that have been illustrated for the case of redox reactions can be applied to other situations including growth of crystals from the vapour phase, in solution, solidification, precipitation and biomineralization reactions to quantitatively predict morphologies in various regimes. Here, we illustrate the applicability of the analysis to the precipitation of hydroxyapatite (Details of calculation are presented in supplementary information). The calculated morphology diagram for hydroxyapatite is illustrated in Fig. 5A. The observed crystal morphologies in this case are also consistent with the predictions of the morphology diagram. Fig. 5B and 5C illustrate the 2-D and 3-D morphologies obtained at low and high driving forces, respectively. The 2-D crystal that is formed under these conditions has the [100] orientation (flat prism plane) and grows along the [002] direction as is seen from the SAD pattern (inset). The prism plane has the lowest surface energy for this structure. It is very interesting to note that the morphology diagram predicts the formation of 2-D crystals under conditions of biomineralization of bone ($37^{o}$C and pH ~ 7.2-7.4). This is identical to the morphology and growth direction of the apatite phase in the bone. At present, to the best of our knowledge, there is no satisfactory explanation for why that apatite phase in the bone is plate-shaped [25]. One of the widely accepted reasons is that the hydroxyapatite phase inherits the shape from the precursor octacalcium phosphate phase. Based on the thermodynamic and kinetic studies, it is shown that OCP is more favourable than the HA in biomineralization conditions and hence the formation of HA happens through the precursor OCP phase. However, we believe that the physiological conditions existing during bone biomineralization (ion concentration, pH and temperature) promote the formation of 2-D structures as evident from Fig. 5B. While the fundamental reason for obtaining 2-D shapes is related to the available thermodynamic force, the presence of biomolecules could provide active control during crystal growth by modifying the kinetics of growth by interactions at the step edges [26].

In literature, there have been considerable efforts directed towards the growth of two-dimensional nanostructures owing to the interesting optical properties that they exhibit. The synthesis protocols used for achieving shape control can be broadly classified as biological, thermal or photochemical. The role of preferential adsorption of surfactants for achieving shape control is over-emphasised in many of these methods. However, it is obvious that preferential adsorption on {111} facets of the growing FCC crystals will lead to the formation of shapes in which all the {111} facets express themselves (octahedron/tetrahedron) and will not lead to the formation of plate-shaped structures. The kinetic control hypothesis predicts that 2-D nanostructures form only under conditions where the reaction is considerably slowed down. The reducing agents that are used for this purpose are typically weak reducing agents. Addition of reagents that favours the backward reaction is seen to have a profound effect on the formation of 2-D structures. In all the cases, it is clearly seen that kinetic control mechanistically corresponds to the regimes with low driving force for reduction of the metal ion and that is responsible for the formation of the 2-D nanostructures. With a detailed knowledge of the reduction potentials of the reducing agents used and the corresponding reactions, it should be possible to develop morphology diagrams for rational synthesis of 2-D nanostructures using any combination of metal salt/reducing agent. Based on extensive analysis of the literature on synthesis of two-dimensional morphologies, we wish to emphasise that the formation in every instance is directly controlled by the driving force and that the surfactants are only incidental for the formation of such structures. Preliminary investigation shows that two-dimensional morphologies can be obtained in several other systems including ZnO and $CaCO_3$ without using any surfactant (Supplementary Information). We believe that the primary role of the surfactant is in providing size control in the formation of the two-dimensional structures. However, the role of surfactant is critical for shape-controlled synthesis of various three dimensional morphologies like cubes[27], rods[28] and wires[29].

The use of radiation has been exploited successfully for producing monodisperse silver nanoprisms [2,3]. We believe that the role of light in photoejection of electrons and fragmentation of the nanoparticles provides a backward reaction that causes a reduction in the net driving force to enable shape-controlled growth by the 2-D nucleation mechanism. Biological synthesis had been employed for obtaining Au platelets, but it is important to note that gold chloride in aqueous medium spontaneously results in the formation of Au platelets. We strongly believe that most of the synthesis conditions( under which platelets are formed) (see supplementary information), have the combination of weak reducing agent, acidic medium, concentration and temperature essentially achieve low driving force that is required for the growth of platelets. The formation of nanobelts and 2-D nanostructures during vapour-phase synthesis can also be rationalised based on the analysis of the driving forces. For many applications, it is desirable to be able to produce ligand-free nanoparticles. For instance, it has been pointed out that the presence of oleic acid and oleyl amine ligands could alter the magnetic properties and electronic configuration of capped FePt nanoparticles and hence a scheme was proposed to remove the ligands after nanoparticle formation. For biomedical applications, it is imperative that the nanostructures are surfactant-free or have surfactants that are biocompatible. In this context, it would be advantageous to be able to produce these structures without the use of surfactants. However, surfactants would still be necessary to obtain size control during synthesis.

**Methods**

The synthesis procedure involved heating aqueous solution of noble metal salts under various conditions (without any reducing/capping agent). The temperature and the pH values were chosen to confirm predictions of the morphology diagram. For low temperature syntheses (below 100$^{o}$C), the reactions were carried out either under reflux conditions or in an autoclave. For temperatures above 100$^{o}$C, the reactions were carried out in a teflon-lined autoclave. The pH of the solutions was controlled using buffer solutions. For neutral pH, 25 ml of 0.1 M $KH_2PO_4$ + 14.55 ml of 0.1 M NaOH mixture was used. For acidic pH between 3 to 6, acetic acid + sodium acetate mixture was used while for acidic pH of 2, 50ml of 0.2 M KCl + 13 ml of 0.2M HCl mixture was used. The efficacy of the buffers at higher temperatures (~ 150$^{o}$C) under hydrothermal conditions is not certain. There could be changes in the pH owing to 'degradation' of the buffer. In some cases, no buffer was used. In these cases, the pH of the solution changes continuously as the reaction proceeds and hence the driving force also changes.

1 mM aqueous solution of $HAuCl_4$ was prepared by dissolving 16 mg of $HAuCl_4$ in 40ml of de-ionized water. Heating the solution to 150$^{o}$C for 4 hours leads to exclusive formation of 2-D gold nanostructures. The nanoplates exhibit triangular, truncated triangular and hexagonal shapes (approximately 45%, 40% and 15%, respectively). Analysis of more than 200 nanoprisms indicate that the lateral dimensions, as measured from visible-light, SEM and TEM images range from 300 nm to 17 μm with a significant fraction exhibiting a size of around 5 μm.

The formation of silver takes place only under basic conditions as it is evident from the morphology diagram. In a typical synthesis, 10 mM aqueous solution of $AgNO_3$ was prepared by dissolving 67 mg of $AgNO_3$ in 40ml of water. Ammonium hydroxide solution was added to raise the pH to 8. Then the reaction mixture was heated in a teflon autoclave at 150°C for 4 hour. 2-D nanostructures of Ag were formed under this condition. For the synthesis of Pt, 1 mM aqueous solution of $H_2PtCl_6$ was prepared by dissolving 20 mg of $H_2PtCl_6$ in 40 ml of acetate buffer solution having a pH of 3. Then, the reaction mixture was heat-treated at 150°C for 4 hours. Pt platelets were formed under this condition. When the reaction is carried out above pH of 4, 3-D porous Pt clusters were formed. Two-dimensional nanostructures of Pt were also obtained when the aqueous solution of $H_2PtCl_6$ was heated in the autoclave at 200°C for 12 hours without the addition of a buffer. Similarly, when 1 mM aqueous solution of $K_2PdCl_6$ was heated in the autoclave at 225°C for 12 hours (without any buffer), Pd nanostructures consisting of triangular and hexagonal platelets were formed.

Hydroxyapatite was synthesized by adding the aqueous solutions of $(NH_4)_2HPO_4$ to an aqueous solution of $Ca(NO_3)_2$. The pH of the solution was maintained at 6 by adding the acetate buffer. This reaction mixture was heated at 150°C for 4 hours. 2-D sheets/films of hydroxyapatite with the brad faces corresponding to the prism plane were obtained as a product. When the synthesis was carried out under pH of 12 by adding concentrated NaOH, 3-D clusters of hydroxyapatite were obtained.

The samples are characterized using a combination of microscopy techniques. SEM was carried out using a FEI Sirion FESEM operated at 5-10 kV. TEM studies were carried out in a JEOL 200CX TEM operated at 160 kV. High-resolution microscopy studies were carried out using a Tecnai F30 field-emission TEM operated at 300 kV.

**Acknowledgements** We thank Dr. Srinivasan Raghavan, Dr. Yamuna Krishnan, Prof. K. Chattopadhyay, Prof. T.A. Abinandanan and Prof Ranganathan for useful comments. Financial assistance from DST NSTI scheme and CSIR is acknowledged. The high-resolution microscopy facilities are a part of the National Facility for Electron Microscopy at the Indian Institute of Science.

* Materials Research Centre, Indian Institute of Science, Bangalore 560012, India.
Fax: +91-80-2360-7316 ; Tel:+91-80- 2293-3255
nravi@mrc.iisc.ernet.in.

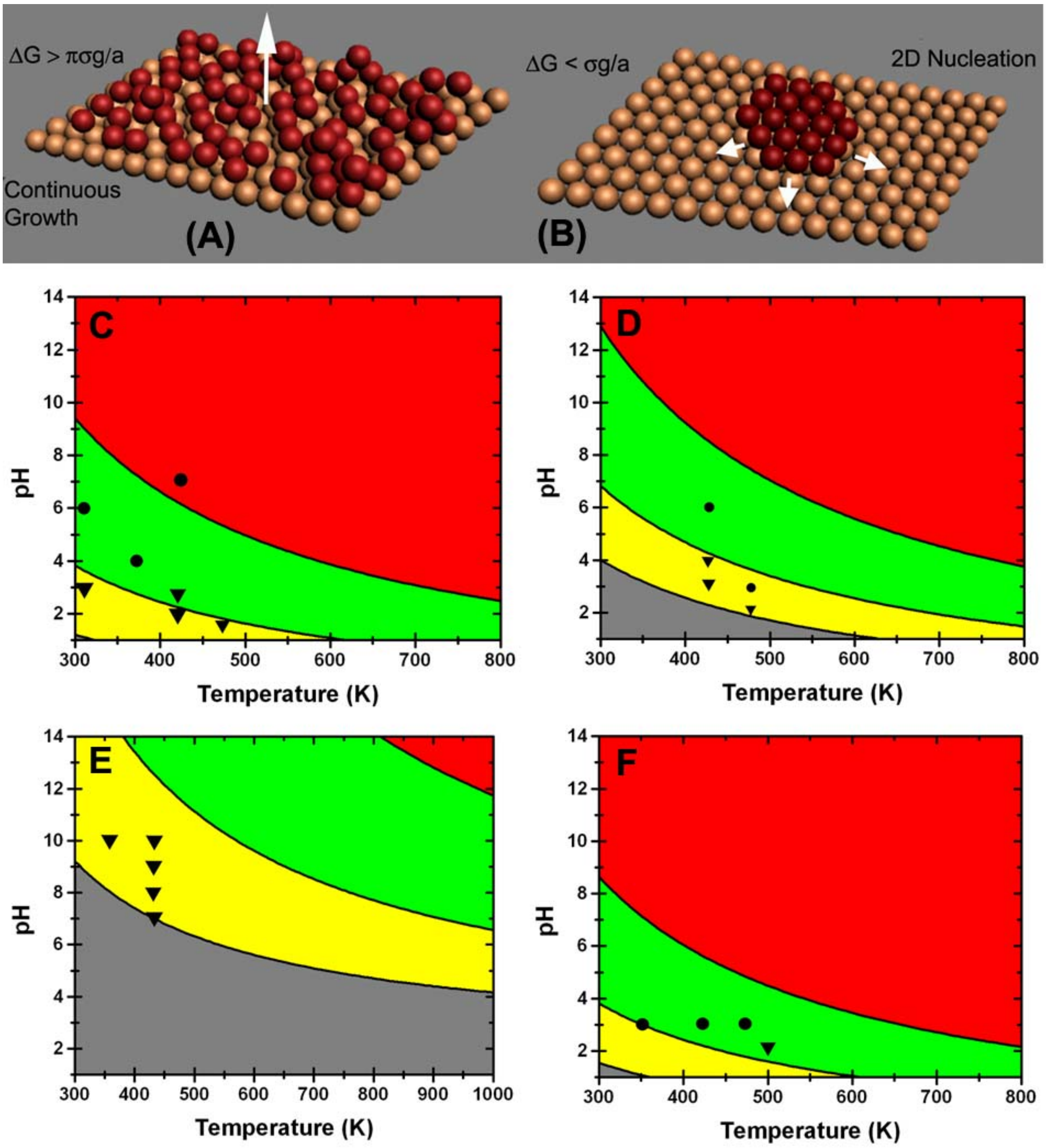

**Fig. 1**. Schematic illustration of **(A)** Continuous growth mechanism and **(B)** Two-dimensional nucleation mechanism. At lower driving forces, the interface cannot move normal to itself and growth has to proceed by lateral motion of steps. At higher driving forces, the interface can move normal to itself, leading to continuous growth. Growth by the two-dimensional nucleation mechanism leads to the formation of structures in which one dimension (thickness) is much smaller than the lateral dimensions. Morphology diagrams illustrating pH and temperature regimes where the two-dimensional nucleation mechanism (yellow) and continuous growth (red) is operative for **(C)** Au, **(D)** Pt, **(E)** Ag and **(F)** Pd. The calculations are based on metal ion concentration of 1 mM for Au, Pt and Pd and 10 mM for Ag. The green region represents the transition regime where there is a gradual transition from 2-D growth at lower temperature/pH towards three-dimensional growth as one approaches the region marked red. Physically, this corresponds to vanishing free energy for step nucleation as one approaches the continuous growth region.

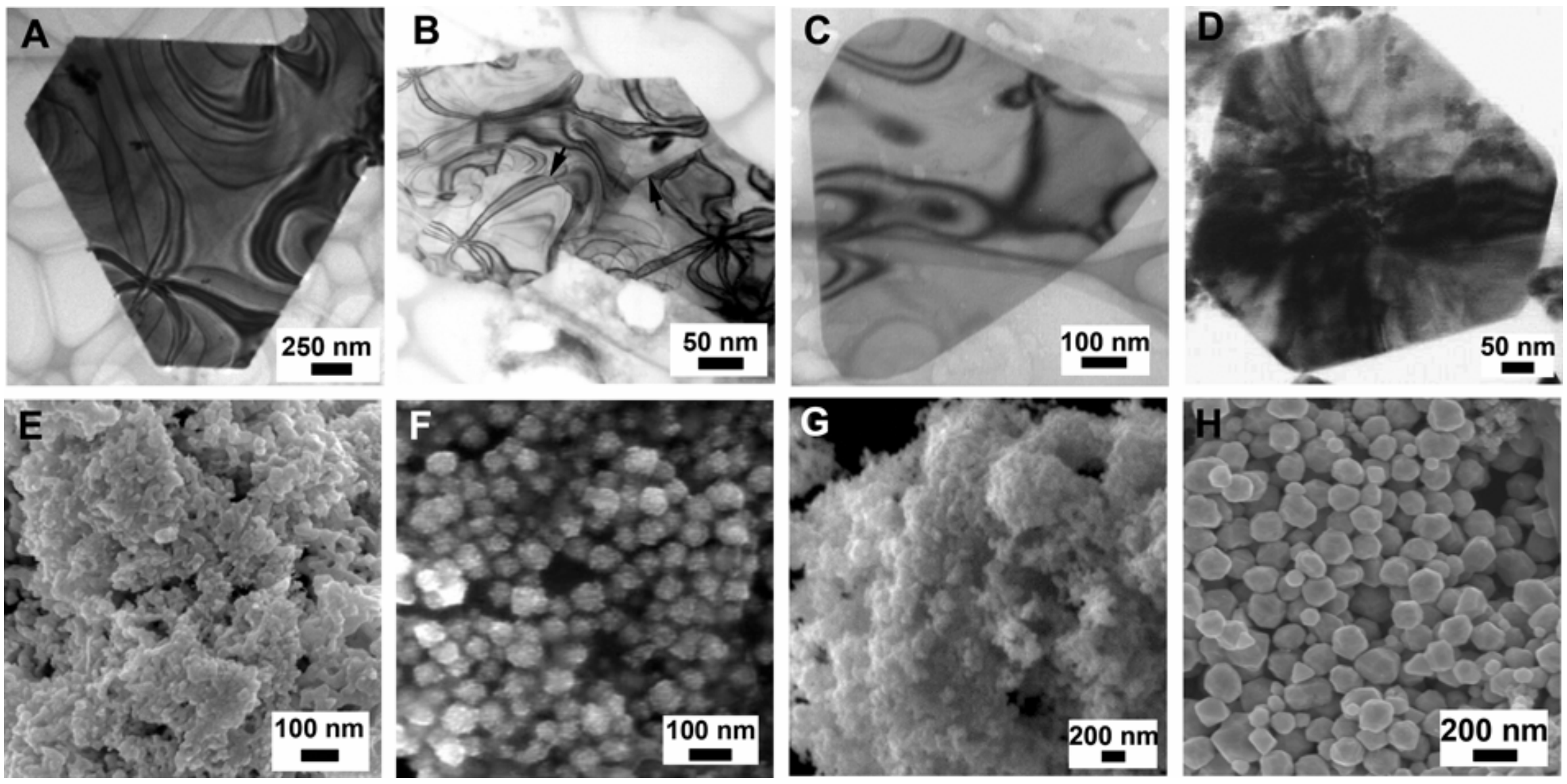

**Fig. 2**. Two-dimensional nanostructures formed under low driving forces for **(A)** Au, **(B)** Pt, **(C)** Ag and **(D)** Pd. Bend contours characteristic of thin platelets are clearly seen. In the case of Pt (B), the discontinuity in the bend contours (arrowed regions) indicates local difference in thickness due to presence of a surface step. Three-dimensional structures formed by the continuous growth mechanism for **(E)** Au, **(F)** Pt, **(G)** Ag and **(H)** Pd.

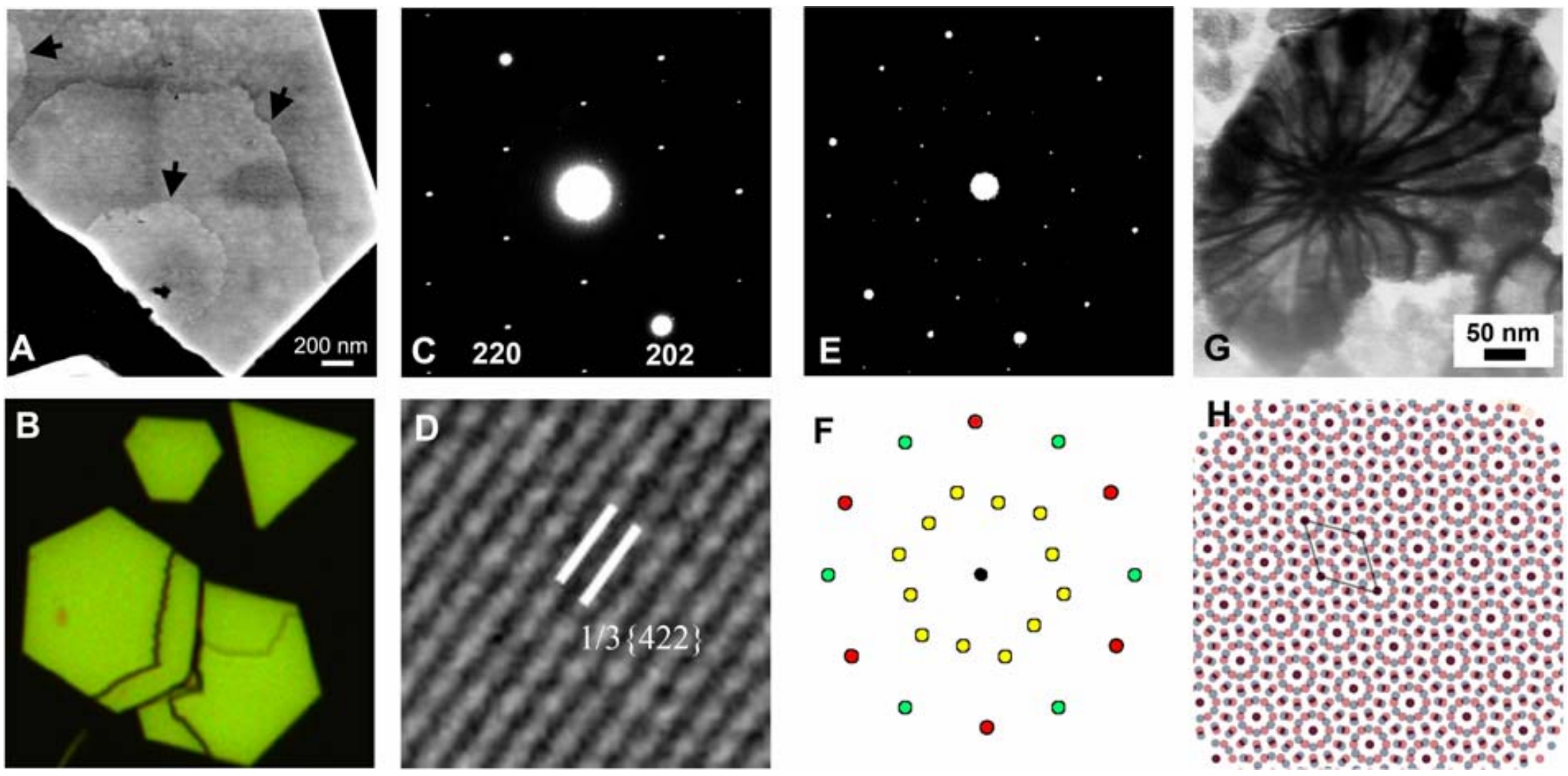

**Fig. 3.** **(A)** Secondary-electron image showing steps on the surface of a Pt platelet (marked by arrows). **(B)** Visible-light microscope image using Nomarski contrast illustrating the presence of steps on a gold platelet. The green color is due to the filter that was used to acquire the image. **(C)** SAED pattern from a silver platelet showing the presence of the kinematically forbidden 1/3{422} reflections due to the presence of stacking faults. **(D)** High-resolution image from a gold platelet showing the fringes corresponding to 1/3{422} lattice spacing. **(E)** SAED pattern from a Pt platelet showing two [111] zone patterns rotated by 27.8$^o$ indicating the presence of a twist boundary. The schematic **(F)** shows the two [111] patterns of Pt (in red and green respectively). The yellow spots that appear closer to the direct beam arise due to double diffraction. **(G)** Bright-field image from the Pt platelet showing the presence of six sets of bend contours extending throughout the platelet indicating that the twist boundary extends completely laterally. **(H)** Atomic structure (un-relaxed) of the $\Sigma$13 boundary that is formed as a result of the rotating one half of the crystal by 27.8$^o$ about a [111] direction.

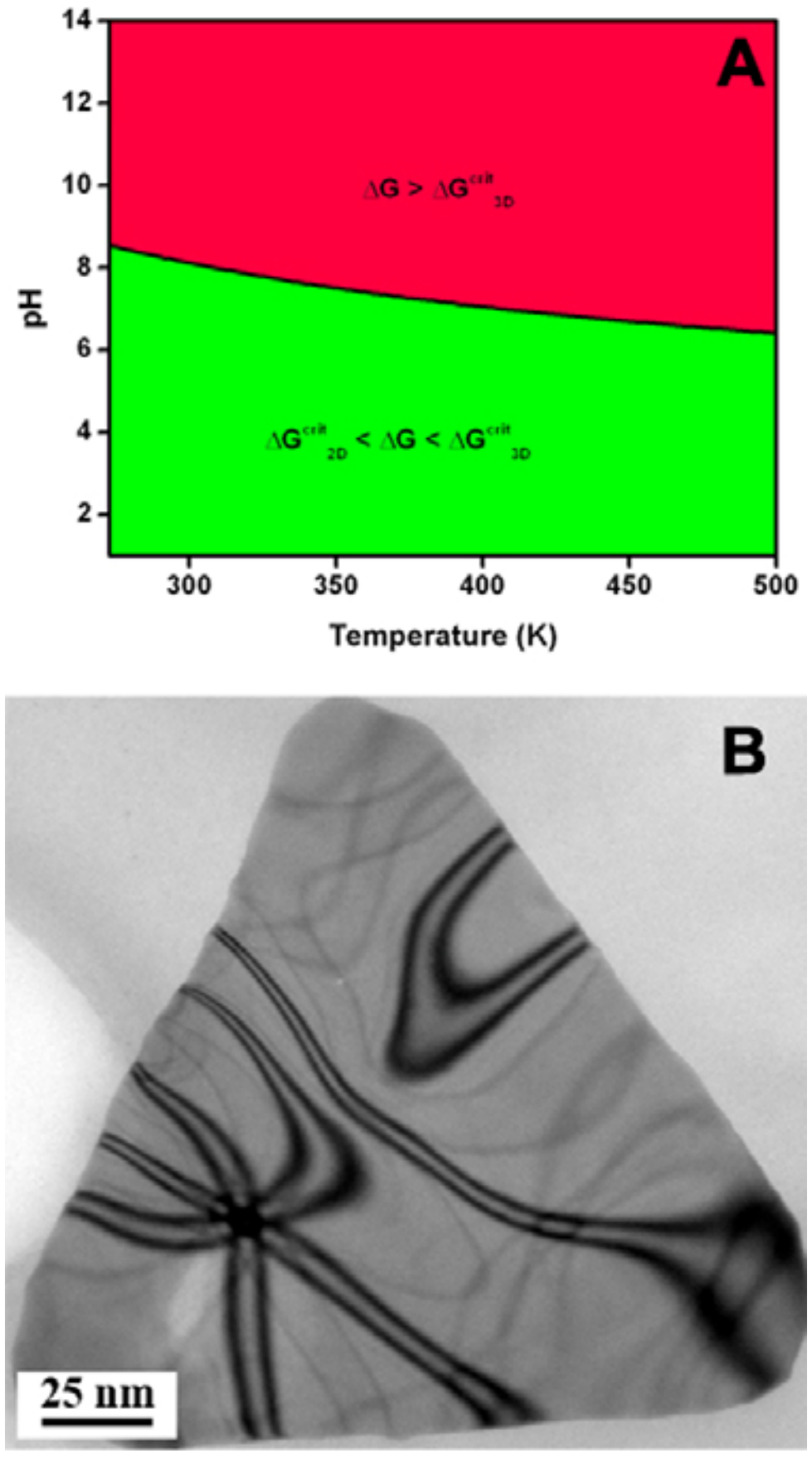

**Fig. 4. (A)** Morphology diagram for the Ag in the presence of the reducing agent $NaBH_4$ showing 2D-3D transition regime and 3D regime. **(B)** TEM bright field image of Ag platelet synthesised at room temperature using $NaBH_4$ in pH=5. The contrast in the image is due to the bend contours.

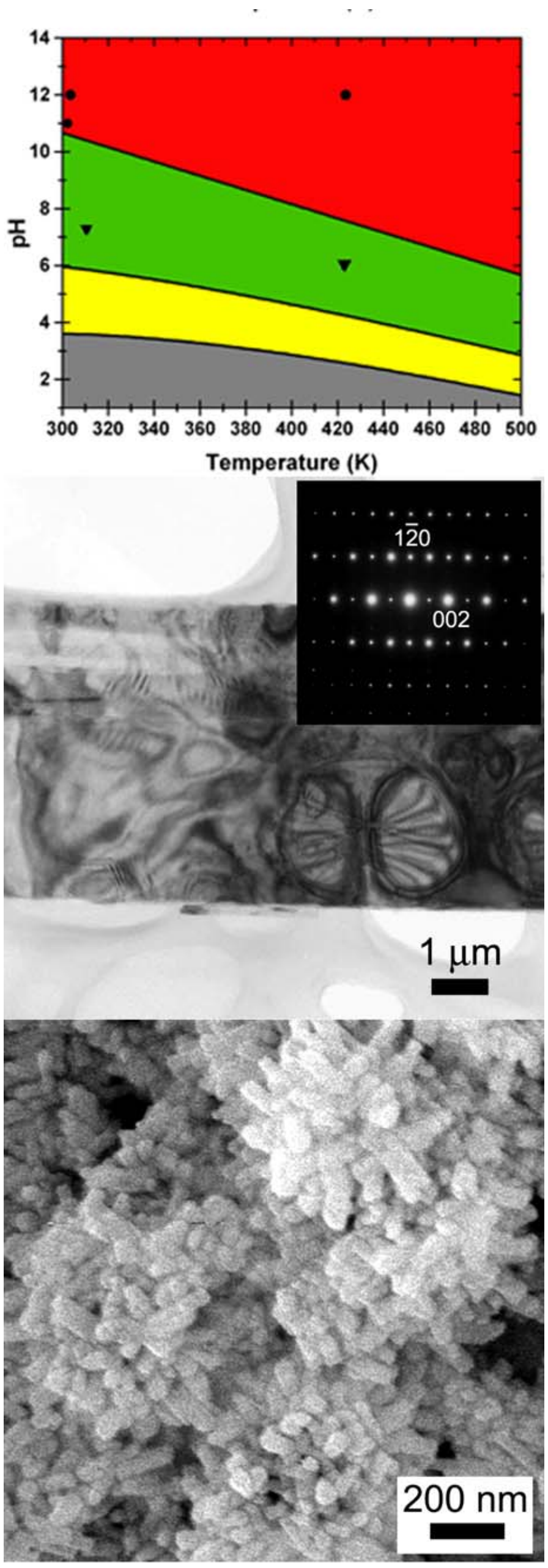

**Fig. 5. (A)** Morphology diagram for the precipitation of hydroxyapatite. The triangles represent experimental data points where 2-D platelets are obtained as predicted by the diagram. The filled circle represents the region where the continuous growth mechanism is operative leading to 3-D structures. **(B)** Bright-field TEM image of plate-shaped apatite phase with the corresponding diffraction pattern inset. The broad face of the plates correspond to the prism planes which grow along the [001] direction. The platelets undergo beam damage on exposure to the electron beam. **(C)** Continuous growth under higher driving force leads to the formation of 3-D structures as seen in this SEM image.

# Predicting Growth of Two-Dimensional Nanostructures

**B. Viswanath, Paromita Kundu, B. Mukherjee and N. Ravishankar***

## Supplementary Information

# I. Calculation of critical driving forces for 2-D and 3-D growth and morphology diagrams

The critical driving forces are calculated based on the expressions derived by Cahn (Ref. 20 and 21, main paper) and are given as

$-\Delta G_{2D} < \sigma/a$ and $-\Delta G_{3D} > \pi\sigma/a$

Thus, 2-D structures are predicted to form for driving force ($-\Delta G$) below $\sigma/a$ while 3-D structures form above a driving force of $\pi\sigma/a$. Here, $\sigma$ is the metal/water interfacial energy and 'a' is the monatomic step height on the growing surface. The interfacial energies and step heights used for different systems are listed below. Reliable data for metal/water interfacial energy is not available. However, contact angle data for water on metals indicates that the value of metal surface energy and metal/water interfacial energy are very close. Thus, the surface energy values of metals are used for the calculations. For the hydroxyapatite/water system, the interfacial energy has been taken from Ref. 1 while the monatomic step height has been taken from Ref. 2.

**Table.T1** Values of interfacial energies and monatomic step heights on the surface that have been used to calculate the critical driving forces for 2-D nucleation and continuous growth

| System | Interfacial Energy $mJ/m^2$ | Monatomic Step Height (Å) | $-\Delta G_{2D}$ kJ/mol | $-\Delta G_{3D}$ kJ/mol |
|---|---|---|---|---|
| Au (111) | 1.39 | 2.3 | 60 | 188 |
| Ag (111) | 1.09 | 2.36 | 46 | 144 |
| Pt (111) | 2.20 | 2.26 | 97 | 304 |
| Pd (111) | 1.74 | 2.24 | 78 | 245 |
| $Ca_5(PO_4)_3(OH)$ (100) plane | 10.4 | 2.76 | 6 | 18.8 |

The actual driving forces under different conditions have been calculated using $E^o$ values and applying the Nernst equation. The calculation for gold is given below. For hydroxyapatite precipitation, the free energy (driving force) has been calculated as $\Delta G = - RT/\nu \ln(IAP/K_{sp})$ where IAP is the ionic activity product of hydroxyapatite and Ksp is the solubility product. The details of the calculations are presented below.

Noble metal salts can be reduced to the corresponding metals by exploiting the oxidation of water. Thus, these metals can be formed without the addition of any external reducing agent. The corresponding reaction can be represented as follows:

$HAuCl_4 + 3e^- \rightarrow Au + 4Cl^- + H^+$
$3/2H_2O \rightarrow 3/4\ O_2 + 3H^+ + 3e^-$

$Ag^+ + e^- \rightarrow Ag$
$1/2H_2O \rightarrow 1/4O_2 + H^+ + e^-$

$H_2PtCl_6 + 4e^- \rightarrow Pt + 6Cl^- + 2H^+$
$2H_2O \rightarrow O_2 + 4H^+ + 4e^-$

$PdCl_6^{2-} + 4e^- \rightarrow Pd + 6Cl^-$
$2H_2O \rightarrow O_2 + 4H^+ + 4e^-$

## Gold:

The details of the free energy calculations for gold are illustrated below.

$HAuCl_4 + 3e^- \rightarrow Au + 4Cl^- + H^+$ ($E^o = 1.002$ V)

$3/2H_2O \rightarrow 3/4\ O_2 + 3H^+ + 3e^-$ ($E^o = -1.229$ V)

-----------------------------------------------------------

$HAuCl_4 + 3/2\ H_2O \rightarrow Au + 4Cl^- + 3/4\ O_2 + 4H^+$ ($E^o = -\ 0.227$)

From Nernst equation, $E = E^o – (RT/nF) \ln K$

Where K is equilibrium constant for the reaction

By substituting $E^o$= -0.227 V, R= 8.314 $JK^{-1}mol^{-1}$, n=3 and F = 96500,

$E = -0.227 – 0.000066138(T) \log\{[Au]\ [Cl^-]^4\ [O_2]^{3/4}\ [H^+]^4 / [HAuCl_4]\ [H_2O]^{3/2}\}$

Activity of Au, $H_2O$ and $O_2$ are 1. Thus,

$E = -0.227 – 0.000066138(T) \{4log[Cl^-] + 4log[H^+] – log[HAuCl_4]\}$

Substituting the concentration of $Cl^-$ in terms of $[Au^{3+}]$,

$E = -0.227 + 0.000066138(T) \{4pH- 3log[Au^{3+}] – 2.4082\}$

Thus, $\Delta G = + 65.7165- 0.019146951(T) \{4pH- 3log[Au^{3+}] – 2.4082\}$

The locus of points where $-\Delta G = 0$, 60 and 188 kJ/mol (from Table T1) delineate different regions in the morphology diagram for gold.

# Hydroxyapatite:

$5Ca(NO_3)_2 + 3(NH_4)_2HPO_4 + 4H_2O \rightarrow$

$Ca_5(PO_4)_3(OH) + 6NH_4^+ + 10NO_3^- + 4H^+ +3H_2O$

Free energy change for precipitation of hydroxyapatite is calculated as follows

$\Delta G = - RT/\nu \ln (IAP/ K_{sp})$

Where, $\nu$ is the total number of ions present in the product. For hydroxyapatite, $\nu = 9$.

The solubility product of hydroxyapatite as a function of temperature (Ref 3) can be expressed by the following equation.

$logK_{sp} = [(- 8219.41/T) – 1.6657 – 0.098215T ]$

Ionic Activity product (IAP) is the product of the activities of the ions present in hydroxyapatite.

$IAP = [Ca^{2+}]^5[PO_4^{3-}]^3[OH^-]$

To determine IAP, we need to calculate the activities of the ions in solution. The activity coefficients are calculated using Davies equation. Ionic strength of the reaction medium is calculated to be 0.137 and hence Davies equation is used to calculate the activity coefficients of the ions.

$\log(\gamma) = -0.511Z^2 \{(I^{1/2} / 1+ I^{1/2})- 0.3I\}$ for $I< 0.5$.

$\log(\gamma) = -0.117 Z^2$ (for $I = 0.137$)

Using the above equation, activity coefficients (γ) of $Ca^{2+}$, $PO_4^{3-}$ and $OH^-$ are found to be 0.34, 0.088 and 0.76 respectively.

Activity (a) = γM

Activity of the $Ca^{2+}$ ion = 0.34(0.001) = $3.4x10^{-3}$

Activity of the $OH^-$ ion = 0.76 [$OH^-$]

Activity of the $PO_4^-$ ion = 0.088 [$PO_4^{3-}$]

The concentration of phosphate ion is calculated by considering the dissociation of $HPO_4^-$ in the solution.

$K(HPO_4^{2-}) = [H+] [PO_4^{3-}] / [HPO_4^{2-}]$

$[PO_4^{3-}] = K_{[HPO42-]} [HPO_4^{2-}]/[H^+]$

By substituting the known concentration of $HPO_4^{3-}$ (0.0056) and its dissociation constant (6.6 X $10^{-13}$), the actual concentration of $PO_4^{3-}$ and its activity is determined.

Activity of the $PO_4^{3-}$ ion = 0.088 {0.036 x $10^{-13}/[H^+]$} = (3.16 X $10^{-16}$) / [$H^+$].

Combining the expressions for activities of the ions,

$IAP = 10.89x10^{-60} \{[OH^-] / [H^+]^3 \}$

where the concentration of $OH^-$ and $H^+$ ions are pH dependent variables.

$IAP = 10.89x10^{-60} \{10^{x-14} / (10^{-x})^3\}$, where x = pH.

After simplification, log(IAP) = 4pH- 72.962

After substituting the values for IAP and Ksp, the free energy function is simplified and can be expressed as a function of temperature and pH as follows.

$\Delta G = - [\ 8.508pH\ (T) - 151.6471(T) + 0.2089\ (T^2) + 17482.68\ ]$

**Table T2:** Reactions and driving force as a function of pH and temperature for various systems

| Reaction | ΔG (kJ/mol) |
|---|---|
| $HAuCl_4 + 3e^- \rightarrow Au + 4Cl^- + H^+$ (1.002 V)<br>$3/2H_2O \rightarrow 3/4O_2 + 3H^+ + 3e^-$ (-1.229 V) | $65.72 - 0.019T\,[4pH - 3\log(C_{Au+3}) - 2.41]$ |
| $Ag^+ + 4e^- \rightarrow Ag$ (0.7996 V)<br>$2H_2O \rightarrow O_2 + 4H^+ + 4e^-$ (-1.229 V) | $41.43 - 0.019T\,[pH + \log(C_{Ag+1})]$ |
| $H_2PtCl_6 + 4e^- \rightarrow Pt + 6Cl^- + 2H^+$ (0.7175 V)<br>$2H_2O \rightarrow O_2 + 4H^+ + 4e^-$ (-1.229 V) | $197.44 - 0.019T[6pH - 5\log(C_{Pt+4}) - 4.67]$ |
| $PdCl_6^{2-} + 4e^- \rightarrow Pd + 6Cl^-$ (0.9395 V)<br>$2H_2O \rightarrow O_2 + 4H^+ + 4e^-$ (-1.229 V) | $112.91 - 0.019T[4pH - 5\log(C_{Pd+4}) - 4.67]$ |
| $5Ca(NO_3)_2 + 3(NH_4)HPO_4 + 4H_2O \rightarrow$<br>$Ca_5(PO_4)_3(OH) + 6NH_4NO_3 + 4HNO_3 + 4H_2O$ | $-8.51pH(T) + 151.65T - 0.21(T^2) - 17482.68$ |

# II. Mechanistic Aspects of Shape Control

## II.1. Role of chloride ions

Halide ions like $Cl^-$ and $Br^-$ are known for their affinity towards the metal surfaces. To eliminate the effect of $Cl^-$ ion on the shape control, we removed all the $Cl^-$ present in the gold chloride solution prior to the reaction by addition of equivalent amount of $AgNO_3$ to the gold chloride solution. The silver chloride was filtered and completely removed from the solution prior to the reaction. The reaction at 150°C for 4 hours yielded Au platelets in this case also. This clearly rules out the role of chloride ions on shape control.

## II.2. Role of light

The role of light on the reduction process has been verified by completely carrying out the reaction in a dark room. This also led to the formation of the two-dimensional Au nanostructures, thus ruling out the role of light/photocatalytic process for shape control.

# III. Characterization

## a) UV-visible spectroscopy studies

In order to understand the formation mechanism of Au platelets, we have carried out time-dependant UV-visible absorption spectra studies. Samples of gold chloride solution heated at 150°C for different times starting from 0 min to 4 hour have been studied. Initial gold chloride solution shows peak at 300 nm up to 30 minutes clearly indicating the absence of gold nanoparticles/seeds in the beginning stage.

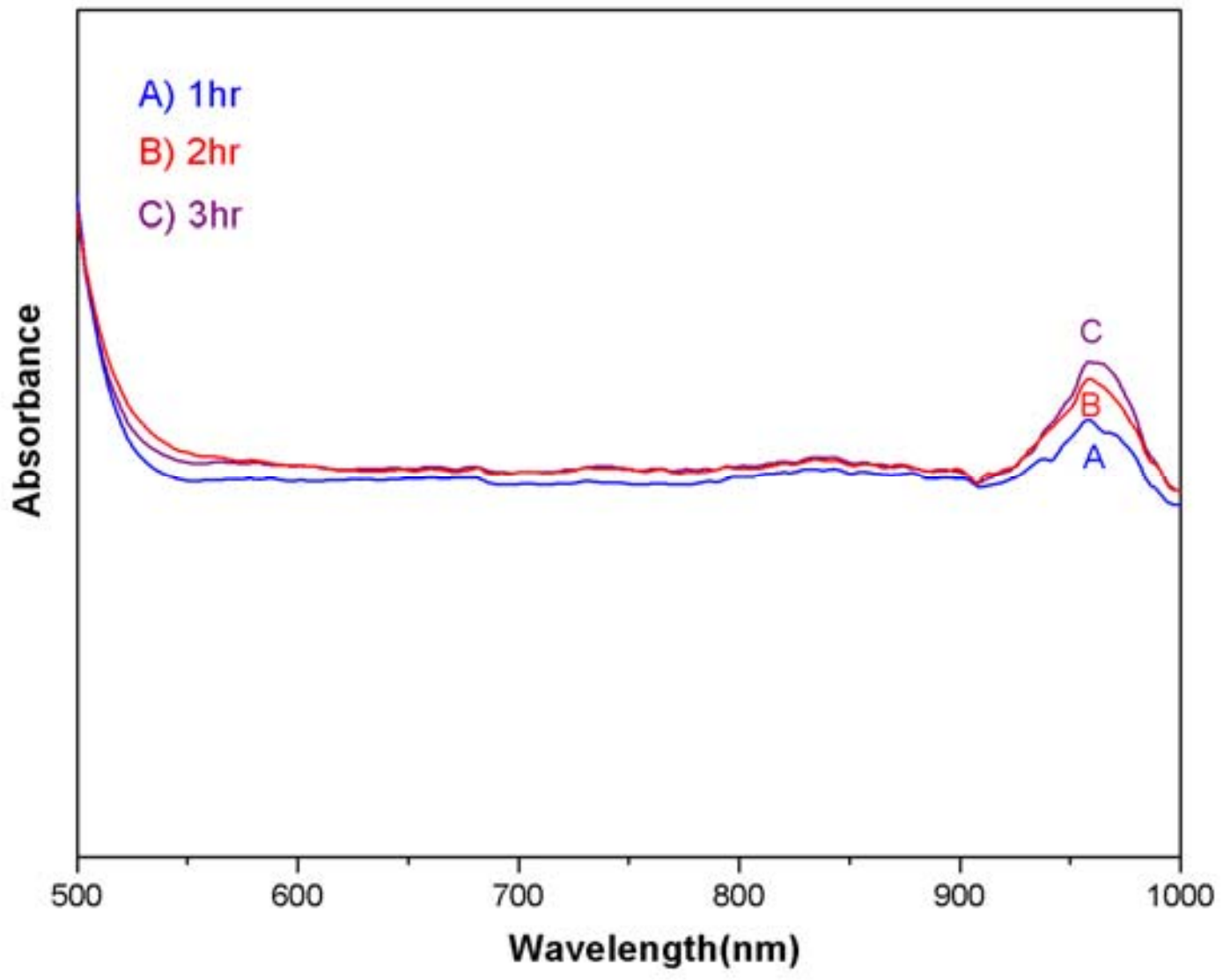

**Fig.S1. UV-visible spectra of Au platelets recorded at different interval of reaction time, shows no peak corresponding to spherical particles and shows only 960nm peak, that corresponds to the Au platelets**

From 1 hour to 4 hour all the solutions show the peak at 960 nm which increases in intensity with time. The peak at 520 nm corresponding to spherical gold nanoparticles/seeds was **not** observed. This strongly suggests that it is not seed-mediated growth of spherical nanoparticles and that the plates form at the earliest stage of the reaction. When we use buffer solution to maintain neutral pH, no Au platelets are observed as was evident from the absence of 960 nm peak in the UV-vis spectrum. UV-vis spectra from silver also show the peak only at higher wavelengths and no peaks corresponding to spherical nanoparticles are observed.

**b) XRD**

To confirm that the 2-D structures are the only product, x-ray diffraction was carried out by collecting the product on a glass substrate. In the XRD pattern for Au, only 111 and 222 peaks are observed clearly indicating that spherical/equiaxed particles are not formed at all. For hydroxyapatite, XRD shows peaks at 100, 200 and 300 that correspond to the prism plane of the hydroxyapatite.

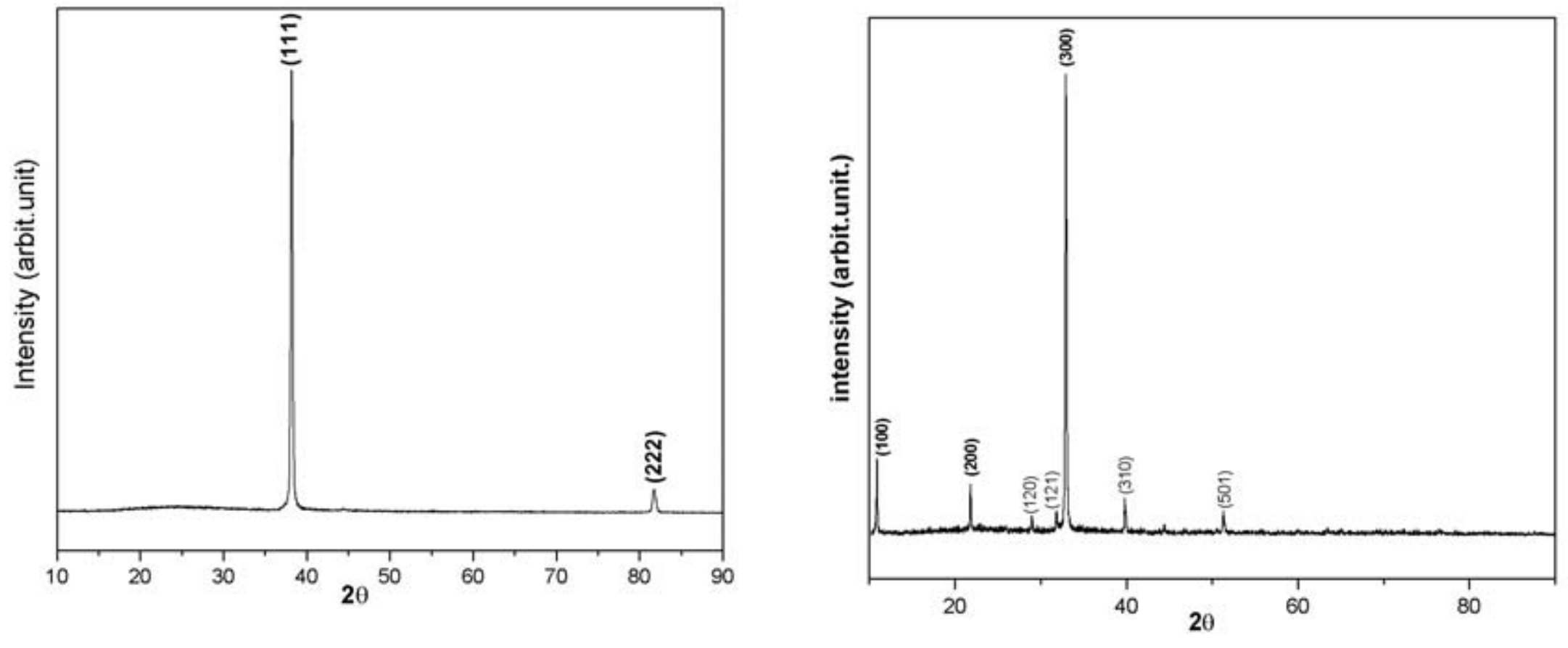

(A) (B)

**Fig. S2** XRD patterns of (A) Au and (B) hydroxyapatite platelets on glass plate. The strong texture due to the (111) planes of Au and (100) planes of hydroxyapatite are evident here.

**c) TEM**

Defects present in these platelets gives strong evidence for the two dimensional growth mechanisms. For instance, platinum shows the steps as well as twist boundary in it that is explained in the text. Because of the presence of steps/twist boundary we see lots of moiré patterns in the bright field image of Pt platelets. This is because of two layers of crystal having the same interplanar distance rotated about certain angle. This is known as rotational moiré pattern in literature.

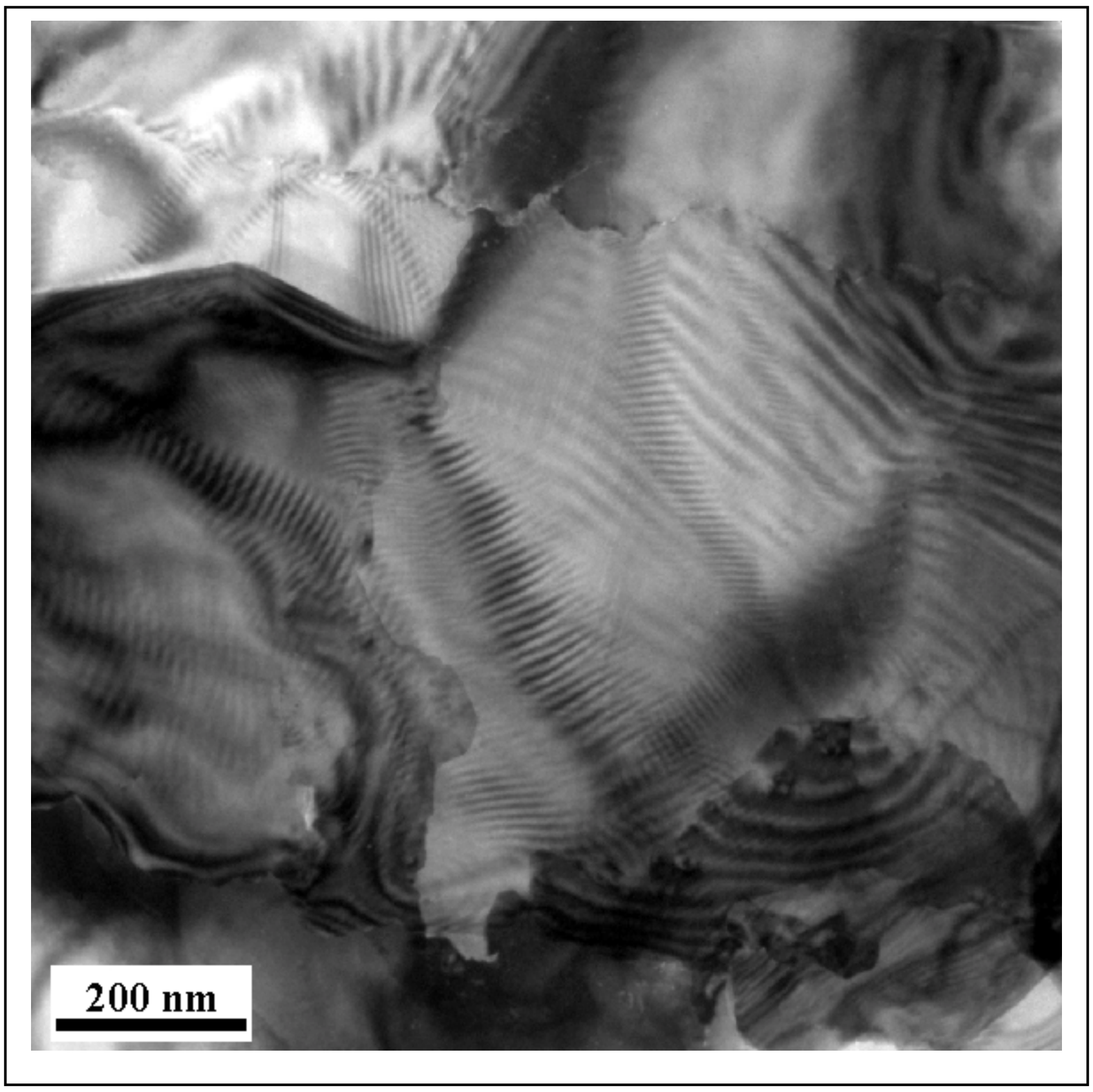

**Figure S3** Bright field image of Pt platelet recorded using transmission electron microscope (TEM) show the presence of moiré fringes and steps.

# IV.Preliminary work on ZnO and $CaCO_3$:

## a) ZnO:

ZnO platelets are synthesized at 200°C keeping the pH=9 using buffer. No surfactants are used in the synthesis. Stability of ZnO is sensitive to pH of the solution, especially at acidic as well as in strong basic conditions, it is not stable. Hence getting complete 3D or 2D is challenging.

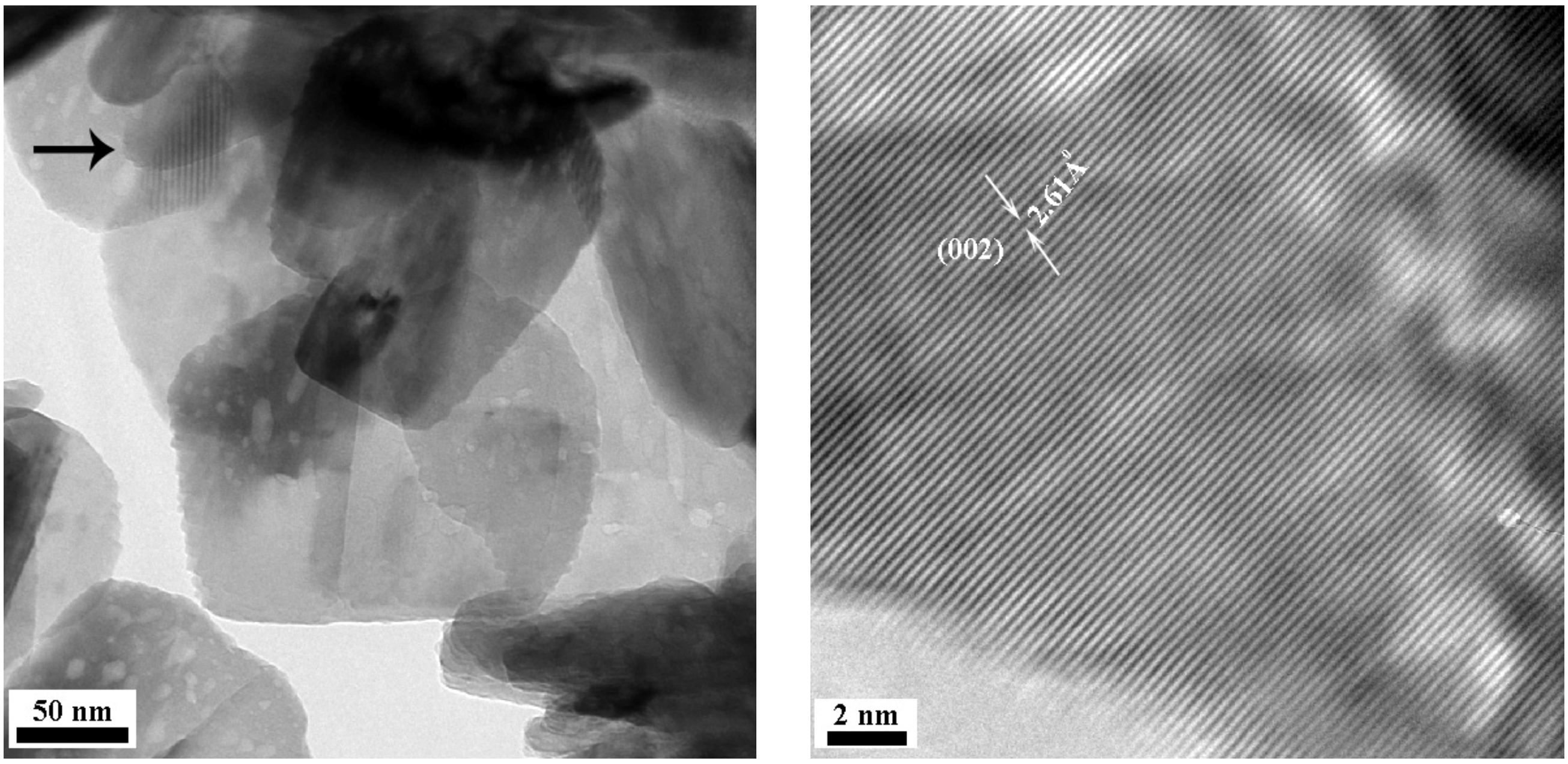

(A) (B)

**Figure S4** (A) Bright field image and (B) lattice image of ZnO platelets recorded using transmission electron microscope (TEM). In figure A, arrow mark shows the presence of moiré fringes that is due to the platelets sitting one over other.

**b) CaCO3:**

Calcium carbonate sheets are synthesised at 200$^{o}$C keeping the pH=6 using buffer. In brief, the aqueous solution of $NaHCO_3$ added into the aqueous solution of $CaCl_2$, then the solution mixture was transferred to Teflon autoclave and heat treated at 200$^{o}$C for 12 hours. No surfactants are used in the synthesis.

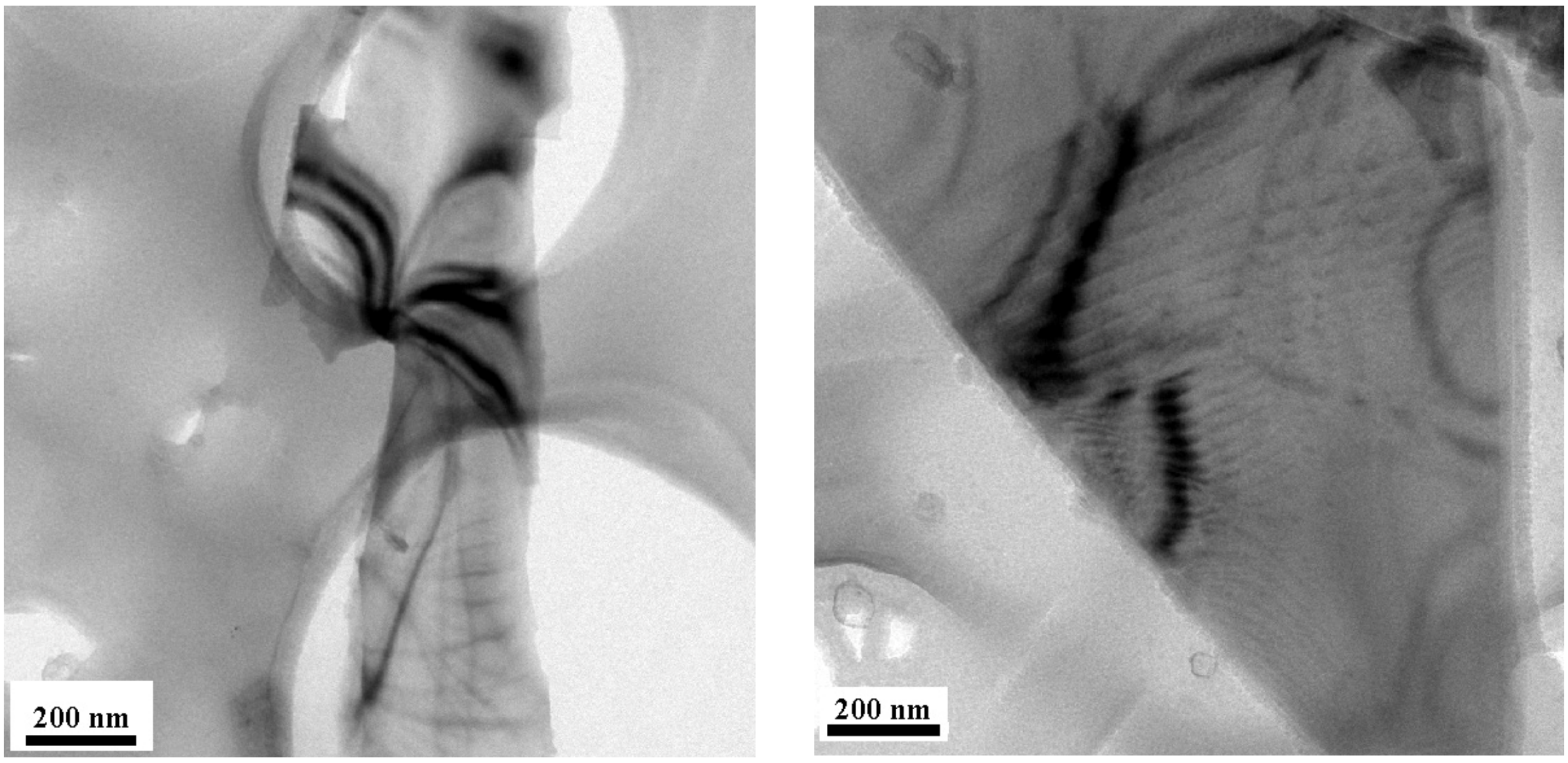

(A) (B)

**Figure S5** (A) Bright field images of CaCO3 platelets recorded using transmission electron microscope (TEM). In fig A, bend contours are visible due to the thin nature of sheets. Fig B shows several steps along with bend contours.